\documentclass[superscriptaddress,aps,prd,nofootinbib,reprint,floatfix]{revtex4-2}

\usepackage{graphicx}
\usepackage{amsmath,amssymb,bm,mathtools}
\usepackage[T1]{fontenc}
\usepackage[colorlinks=true,allcolors=blue]{hyperref}
\usepackage{microtype}
\usepackage{booktabs}

\graphicspath{{figures/}}

\newcommand{\eps}{\varepsilon}
\newcommand{\mpl}{M_{\rm Pl}}
\newcommand{\Br}{\operatorname{Br}}
\newcommand{\Zfour}{\mathbb Z_4}
\newcommand{\ZfourN}{\mathbb Z_{4n}^{\rm(app)}}
\newcommand{\Order}{\mathcal O}
\newcommand{\CP}{\mathrm{CP}}
\newcommand{\dd}{\mathrm d}

\begin{document}

\preprint{TU-1314}

\title{Nelson--Barr  Inflation}

\author{Fuminobu Takahashi}
\email{fumi@tohoku.ac.jp}
\affiliation{Department of Physics, Tohoku University, Sendai, Miyagi 980-8578, Japan}
\affiliation{Kavli IPMU (WPI), UTIAS, University of Tokyo, Kashiwa 277-8583, Japan}

\begin{abstract}
The Nelson--Barr mechanism solves the strong CP problem through spontaneous CP violation, but the resulting CP-conjugate vacua generically lead to stable domain walls. We propose that the Nelson--Barr scalar itself drives hilltop inflation. A CP-symmetric double-well potential for its imaginary component forms two CP-breaking valleys, with inflation proceeding along one of them. Since the lowest CP-invariant ridge is generically much higher than the energy available after inflation, transitions between the two branches are energetically inaccessible, and the domain walls cannot be regenerated after inflation.
A small CP-preserving linear deformation raises the scalar spectral index of quartic hilltop inflation and yields CMB-compatible parameter regions. The Nelson--Barr sector necessarily provides a portal to visible-sector reheating. When coupled to right-handed neutrinos, it can also transmit the spontaneous CP phase to the lepton sector and, in an extension that allows a higher CP-breaking scale, even realize nonthermal leptogenesis.
This establishes a unified cosmological realization of the Nelson--Barr mechanism in which the same CP-breaking dynamics solves the strong CP problem, can generate quark and lepton CP violation, drives inflation and reheating, and enables baryogenesis, while simultaneously eliminating the associated domain-wall problem.
\end{abstract}

\maketitle

\section{Introduction}
\label{sec:introduction}

The physical QCD vacuum angle,
\begin{equation}
 \bar\theta=\theta_{\rm QCD}+\arg\det(M_uM_d),
\end{equation}
is constrained to be extremely small by neutron electric-dipole-moment searches~\cite{Abel:2020gbr}; see Refs.~\cite{Cheng:1987gp,Pospelov:2005pr,Kim:2008hd,Hook:2018dlk} for reviews. The Nelson--Barr mechanism accounts for this small number by imposing CP in the fundamental theory and breaking it spontaneously in a scalar sector coupled to vector-like quarks~\cite{Nelson:1983zb,Nelson:1984hg,Barr:1984qx,Barr:1984fh}. Its characteristic block-triangular quark mass matrix allows an order-one Cabibbo--Kobayashi--Maskawa (CKM) phase while keeping $\bar\theta=0$ at tree level. A minimal realization of this structure is provided by the Bento--Branco--Parada (BBP) model, which introduces a complex singlet scalar and a vector-like down-type quark~\cite{Bento:1991ez}.

Nelson--Barr models face two generic challenges, both of which become particularly important when the spontaneous-CP-breaking scale is high. Higher-dimensional operators and radiative effects can spoil the triangular structure and regenerate $\bar\theta$, leading to the Nelson--Barr quality problem~\cite{Dine:2015jga,Vecchi:2014hpa,Valenti:2021rdu}.
Several ultraviolet mechanisms have been proposed to
improve the quality, including composite or confining
realizations~\cite{Perez:2020dbw,Valenti:2021xjp,Girmohanta:2022giy},
continuous chiral gauge symmetries that forbid the
leading dangerous operators~\cite{Asadi:2022vys,Perez:2023zin},
and approximate discrete
symmetries~\cite{Murai:2024alz}.
Spontaneous CP breaking also produces CP-conjugate
vacua and hence domain walls~\cite{Kobsarev:1974hf}.
If CP is an exact gauge symmetry, the corresponding
walls are exactly stable~\cite{McNamara:2022lrw},
leading to strong cosmological constraints on
Nelson--Barr models~\cite{Asadi:2022vys}.
An economical way to avoid this problem is for inflation
to select a single CP-breaking branch and for only that branch only
to remain populated throughout the subsequent cosmological
evolution. This requires, in particular, that CP not be
thermally restored after inflation~\cite{Murai:2024alz}
and that the post-inflationary dynamics not repopulate the
CP-conjugate branch.

We realize this possibility by identifying the Nelson--Barr scalar itself with the inflaton. Its imaginary component has two CP-conjugate minima that define two inflationary valleys, along each of which the inflaton potential takes a hilltop form. Inflation along one valley homogenizes the CP phase and removes any preexisting walls. This hilltop structure also naturally protects the selected branch after inflation: every continuous path to the conjugate valley must cross the CP-invariant ridge, whose lowest point lies well above the total scalar energy available after inflation. This energetic hierarchy follows from the shallow inflationary plateau together with a CP-odd symmetry-breaking scale that is not parametrically small, as is also relevant for an unsuppressed Nelson--Barr phase.

In the quartic hilltop limit, however, the predicted spectral tilt lies below the observed value. In fact, a small CP-preserving linear deformation moves the CMB pivot closer to the hilltop and reduces the magnitude of the potential curvature sampled by the CMB modes, thereby increasing $n_s$~\cite{Takahashi:2013cxa}. We treat its coefficient as a spurion carrying the same discrete charges as the Nelson--Barr scalar and solve the inflaton evolution numerically to identify CMB-normalized parameter regions.

A successful inflationary cosmology must efficiently populate the visible sector, a requirement that in many models necessitates introducing additional inflaton couplings solely for reheating. In the present setup, however, this connection arises naturally: identifying the inflaton with the Nelson--Barr scalar directly couples it to the Standard Model (SM) through the vector-like-quark sector. This permits tree-level decays involving vector-like and light quarks when kinematically allowed, while the field-dependent heavy-quark threshold induces a loop-level decay into gluons. If right-handed neutrinos are included, the same spontaneous CP phase can also be transmitted to the light-neutrino mass matrix,  accounting for leptonic CP violation. Moreover, in an extension that allows the CP-breaking scale to be raised to around $10^{16},\mathrm{GeV}$, inflaton decays into right-handed neutrinos may generate the baryon asymmetry through nonthermal leptogenesis~\cite{Lazarides:1990huy,Asaka:1999yd,Asaka:1999jb}. Thus, visible-sector reheating and quark-sector CP violation arise naturally within the Nelson--Barr framework, while an appropriate high-scale extension may further connect the same origin of spontaneous CP breaking to leptonic CP violation and baryogenesis.

We finally note the related scenario of Ref.~\cite{Suematsu:2022vqd}, where three real scalar degrees of freedom form a CP-violating valley and realize large-field inflation through nonminimal couplings to gravity. In one of the cases considered there, the CP-breaking order parameter can pass through zero during the post-inflationary oscillation, so the absence of domain-wall regeneration is not manifest from the homogeneous evolution alone. By contrast, our model realizes effectively single-field hilltop inflation along a fixed CP-breaking branch, and the post-inflationary scalar energy is insufficient to cross the CP-invariant ridge.

The paper is organized as follows.  Section~\ref{sec:model} introduces the high-quality Nelson--Barr sector and the CP-preserving linear spurion. Section~\ref{sec:inflation} develops the hilltop inflation scenario and presents a numerical analysis of its inflationary predictions.
Section~\ref{sec:nonrestoration} establishes post-inflationary CP non-restoration.  Reheating and nonthermal leptogenesis are discussed in Sec.~\ref{sec:reheating}.  Section~\ref{sec:conclusion} contains our conclusions.

\section{High-quality Nelson--Barr sector}
\label{sec:model}

\subsection{Discrete charges and the quark mass matrix}

We use the discrete charge assignment for the BBP model proposed in
Ref.~\cite{Murai:2024alz}, which suppresses the leading operators
responsible for the Nelson--Barr quality problem.  Charges are additive
modulo $4$ and $4n$, and the approximate $\ZfourN$ breaking is described
by the complex spurion $X$ introduced there.\footnote{A dynamical realization of the spurion $X$, whose angular mode can
serve as dark matter, was studied in Ref.~\cite{Murai:2024bjy}.}  In addition, we introduce
a second complex spurion $\mathcal C$, carrying the same discrete charges
as the Nelson--Barr scalar, in order to generate the CP-preserving linear
deformation of the inflaton potential.  Their assignments are listed in
Table~\ref{tab:charges}.

\begin{table*}[t!]
\caption{Discrete charges of the fields and spurions.
A barred fermion carries the opposite additive charge.
The spurion backgrounds are chosen real and therefore
preserve CP.}
\label{tab:charges}
\begin{ruledtabular}
\begin{tabular}{c|cccccccccccc}
 & $S$ & $D_L$ & $D_R$ & $Q_{Li}$ & $d_{Ri}$ & $u_{Ri}$
 & $N_i$ & $L_\alpha$ & $e_{R\alpha}$ & $H$ & $X$ & $\mathcal C$ \\
\hline
$\Zfour$
 & $2$ & $2$ & $2$ & $0$ & $0$ & $0$
 & $1$ & $1$ & $1$ & $0$ & $0$ & $2$ \\
$\ZfourN$
 & $2n$ & $2n$ & $2n-k$ & $k$ & $k$ & $k$
 & $n$ & $n$ & $n$ & $0$ & $1$ & $2n$
\end{tabular}
\end{ruledtabular}
\end{table*}

The relevant quark interactions are
\begin{align}
 -\mathcal L_q={}&
 \left(\frac{X}{\Lambda}\right)^k M\,\overline D_LD_R
 +\left(\frac{X^*}{\Lambda}\right)^k
 (g_iS+g_i'S^*)\overline D_Ld_{Ri}\nonumber \\
 &+y^d_{ij}H\overline Q_{Li}d_{Rj}+\mathrm{h.c.},
\label{eq:LNB}
%\end{split}
\end{align}
where $M$, $g_i$, $g_i'$, and $y^d_{ij}$ are real, $D_L$ and $D_R$ are the left- and right-handed components of the vector-like down-type quark, $S$ is the Nelson--Barr scalar, $Q_{Li}$ and $d_{Rj}$ are the SM left-handed quark doublets and right-handed down-type quarks, respectively, $H$ is the SM Higgs doublet, and $X$ is the spurion.

We decompose the Nelson--Barr scalar $S$ into its CP-even and CP-odd components as
\begin{equation}
 S=\frac{s+ia}{\sqrt2},
\label{eq:Sdecomp}
\end{equation}
where $s$ and $a$ transform under CP as  
\begin{align}
    \CP:\quad s\to s,\quad a\to-a.
\end{align}
The scalar develops one of two CP-conjugate vacuum expectation values (vevs),
\begin{equation}
\langle S\rangle_\sigma=\frac{v+i\sigma f}{\sqrt2},
\label{eq:Svac}
\end{equation}
where $\sigma=\pm1$ labels the two minima. Equivalently, $\langle s\rangle=v$ and $\langle a\rangle=\sigma f$.
For generic real flavor couplings, an unsuppressed  CP phase requires $f \sim v$, as we will see shortly.

The approximate $\ZfourN$ symmetry is broken by a real
spurion vev~\cite{Murai:2024alz},
\begin{equation}
 \frac{\langle X\rangle}{\Lambda}
 =\eps\ll1
\label{eq:spurion-vevs}
\end{equation}
with 
\begin{align}
    \varepsilon^k \lesssim 10^{-5}.
    \label{eq:epsilon}
\end{align}
With the charge assignment in Table~\ref{tab:charges},
the vector-like mass term and the $S$-mediated mixing
term in Eq.~\eqref{eq:LNB} both arise at order
$\eps^k$, whereas operators that would spoil the
Nelson--Barr structure require additional powers of
$\eps$.  The quality problem is therefore ameliorated
without suppressing the relative strength of the mass
and mixing terms responsible for the CKM phase~\cite{Murai:2024alz}.

In the vacuum labeled by $\sigma$, the down-sector mass
matrix is
\begin{equation}
 \mathcal M_d^{(\sigma)}
 =
 \begin{pmatrix}
  m_{ij} & 0\\
  B_j^{(\sigma)} & \widehat M
 \end{pmatrix},
\label{eq:massmatrix}
\end{equation}
where
\begin{align}
 m_{ij} &= y^d_{ij}v_H, \\
 B_i^{(\sigma)}
 &=
 \frac{\epsilon^k}{\sqrt{2}}
 \left[
  (g_i+g_i')v
  +i\sigma(g_i-g_i')f
 \right],
 \label{eq:Bvac}\\
 \widehat M &= \epsilon^k M. 
\end{align}
Here $v_H$ denotes the vev of the
SM Higgs field.  Since $m$ and $\widehat M$ are real,
the block-triangular form of Eq.~\eqref{eq:massmatrix}
implies
\begin{equation}
 \det\mathcal M_d^{(\sigma)}
 =
 \widehat M\det m\in\mathbb R.
\end{equation}
The spontaneous phase contained in $B_i^{(\sigma)}$
therefore does not contribute to $\bar\theta$ at tree
level.

The same phase does, however, enter the left-handed
quark mixing.  Suppressing the vacuum label $\sigma$, the left-handed rotation relevant for the
CKM matrix is obtained by diagonalizing
$\mathcal M_d\mathcal M_d^\dagger$.  In the basis
$(d_{Li},D_L)$,
\begin{equation}
 \mathcal M_d\mathcal M_d^\dagger
 =
 \begin{pmatrix}
  mm^T & mB^\dagger\\
  Bm^T & m_D^2
 \end{pmatrix},
 \qquad
 m_D^2=\widehat M^2+|B|^2.
\end{equation}
Integrating out the heavy left-handed state to leading
order in $m/m_D$ gives the effective mass-squared matrix
for the light left-handed down-type quarks,
\begin{equation}
 m\left(
  \bm 1-\frac{B^\dagger B}{m_D^2}
 \right)m^T.
\label{eq:Hlight}
\end{equation}
Although $m$ is real, $B^\dagger B$ is a complex
Hermitian matrix whenever the components of $B$ carry
nontrivial relative phases.  These phases consequently
enter the unitary matrix that diagonalizes
the above mass-squared matrix and can generate the CKM phase.

Writing
\begin{equation}
 B_i^{(\sigma)}=|B_i|e^{i\phi_i},
\end{equation}
the relative phase satisfies
\begin{equation}
 \sin(\phi_j-\phi_i)
 =
 \frac{
  \operatorname{Im}\!\left[
   (B_i^{(\sigma)})^*B_j^{(\sigma)}
  \right]
 }{|B_i||B_j|},
\label{eq:relative-phase}
\end{equation}
where
\begin{align}
 &\operatorname{Im}\!\left[
  (B_i^{(\sigma)})^*B_j^{(\sigma)}
 \right]
 =
 \sigma\epsilon^{2k}vf
 \left(g_i'g_j-g_ig_j'\right),
 \label{eq:ImBB}\\
 &|B_i|^2
 =
 \frac{\epsilon^{2k}}{2}
 \left[
  v^2(g_i+g_i')^2
  +f^2(g_i-g_i')^2
 \right].
 \label{eq:Bnorm}
\end{align}
For generic untuned couplings, an order-one relative phase naturally
requires $f\sim v$ and no accidental alignment between $g_i$ and
$g_i'$ in flavor space.

Finally, for these relative phases to have an
unsuppressed effect on the light-quark mixing matrix, the
correction in Eq.~\eqref{eq:Hlight} must not be small.
This requires
\begin{equation}
 \frac{|B|^2}{\widehat M^2+|B|^2}=\Order(1).
 \label{eq:unsuppressed-mixing}
\end{equation}
Both $B_i$ and $\widehat M$ carry the same overall
factor $\epsilon^k$.  This factor therefore cancels from
Eq.~\eqref{eq:unsuppressed-mixing}, so the approximate
discrete symmetry suppresses dangerous corrections
without suppressing the CKM phase.  This is the essential
advantage of the spurion construction of
Ref.~\cite{Murai:2024alz}.

Lastly, we note a potential quality problem associated with the fact that $S^2$ is neutral under both discrete symmetries.\footnote{We thank Kazunori Nakayama and Kai Murai for pointing out this upper bound on the CP-breaking scale. 
} Consequently, any operator allowed by the symmetries may be multiplied by powers of $S^2/\Lambda^2$. In particular, the following CP-invariant higher-dimensional operator is generically allowed:
\begin{align}
\frac{\operatorname{Im}(S^2)}{\Lambda^2}
\frac{\alpha_s}{8\pi}
G^a_{\mu\nu}\widetilde G^{a\mu\nu},
\end{align}
where  $G^a_{\mu\nu}$ is the gluon field strength, and 
\begin{equation}
\widetilde G^{a\mu\nu}
\equiv
\frac{1}{2}\epsilon^{\mu\nu\rho\sigma}
G^a_{\rho\sigma}.
\end{equation}
After spontaneous CP breaking, this operator induces a contribution of order $f^2/\Lambda^2$ to the effective strong CP phase, up to an order-one factor determined by the vacuum phase. Requiring this contribution to satisfy the experimental bound on $\bar\theta$ therefore gives
\begin{equation}
\label{eq:quality}
\frac{f}{\Lambda}\lesssim 10^{-5}.
\end{equation}
This sets an upper bound on the inflation scale, as we will see below.

\subsection{Linear spurion and the strong-CP phase}

We next introduce an independent dimension-three spurion
$\mathcal C$ that generates a linear deformation of the
inflaton potential.  It carries the same discrete charges
as $S$.  Since these charges are self-conjugate, both
$\mathcal C S$ and $\mathcal C^*S$ are allowed.  For a
real spurion background, however, their contributions are
equivalent and can be absorbed into a single real
coefficient.  We therefore write
\begin{equation}
 \Delta V_C
 =
 -\frac{1}{\sqrt2}
 \left(\mathcal C S+\mathrm{h.c.}\right).
\label{eq:Cspurion-general}
\end{equation}
 For the CP-preserving background
\begin{equation}
 \langle\mathcal C\rangle=C\in\mathbb R,
\end{equation}
this reduces to
\begin{equation}
 \Delta V_C
 \longrightarrow
 -Cs.
\label{eq:Cspurion}
\end{equation}

Since $\mathcal C/\Lambda^2$ transforms in the same way
as $S$, insertions of $S$ in higher-dimensional operators
can be replaced by $\mathcal C/\Lambda^2$ without changing
their discrete charges.  The leading induced operators
relevant to the down-quark mass matrix are
\begin{align}
 -\Delta\mathcal L_C={}&
 c_F^i\frac{\mathcal C}{\Lambda^3}
 \left(\frac{X}{\Lambda}\right)^{2k}
 H\overline Q_{Li}D_R
 \nonumber\\
 &+
 \frac{\mathcal C}{\Lambda^3}
 \left(\frac{X}{\Lambda}\right)^k
 \left(c_M S+c_M'S^*\right)
 \overline D_LD_R
 \nonumber\\
 &+
 c_B^i\frac{\mathcal C}{\Lambda^2}
 \left(\frac{X^*}{\Lambda}\right)^k
 \overline D_Ld_{Ri}
 +\mathrm{h.c.}+\cdots ,
\label{eq:Coperators}
\end{align}
where the Wilson coefficients are real at the
CP-symmetric cutoff scale.  The last operator produces
only a real correction to the lower-left block $B_i$ and
therefore preserves the block-triangular determinant.
The first operator fills the upper-right block, denoted by
$F$, while the second induces a generally complex
correction $\delta\widehat M$ to the lower-right block
through the complex vacuum expectation value of $S$.

For small corrections, the resulting shift of the strong
CP phase is
\begin{equation}
 \delta\bar\theta_C
 \simeq
 \operatorname{Im}\left[
  \frac{\delta\widehat M}{\widehat M}
  -\frac{Bm^{-1}F}{\widehat M}
 \right].
\label{eq:theta-block}
\end{equation}
Using $\widehat M=\eps^kM$ and
$|B|/\widehat M=\Order(1)$, the spurion counting gives
\begin{equation}
 |\delta\bar\theta_C|
 \lesssim
\frac{C}{\Lambda^3}
 \left[
  \frac{f}{M}
  +\eps^{2k}\frac{v_H}{m_d}
 \right],
\label{eq:theta-C-bound}
\end{equation}
where $m_d$ schematically denotes the light-quark mass
scale controlling the largest entries of $m^{-1}$, and we used $|B| \sim \widehat M$ and $c_F^i \sim c_B^i = \Order(1)$.
The first term arises from $\delta\widehat M$, whereas
the second comes from the induced upper-right block $F$.

In the inflationary parameter region considered below,
$C/\Lambda^3\ll 10^{-10}$. Although the spurion
background does not introduce an independent complex
phase, its interference with the spontaneous phase of
$S$ can generate an additional contribution to
$\bar\theta$. This contribution is, however, suppressed
by $C/\Lambda^3$ and remains far below the experimental
bound.

\section{Hilltop inflation}
\label{sec:inflation}

\subsection{Potential and vacuum parametrization}

We now turn to the scalar dynamics of the Nelson--Barr
field $S$.  For the charge assignments
introduced above, $S^2$ is a singlet under both discrete
symmetries. In the absence of spurion insertions, the
scalar potential is therefore invariant under
$(s,a)\to(-s,-a)$. Together with CP invariance, this
implies that the potential contains only even powers of
both $s$ and $a$, although mixed terms  are generally allowed. Odd powers of $s$
may be generated by insertions of the real spurion $\mathcal C$.

To exhibit the inflationary dynamics and the separation
between the two CP-conjugate branches as simply as
possible, we consider the separable potential\footnote{The Coleman--Weinberg correction from the heavy
vector-like quark mainly renormalizes the coefficients in
Eq.~\eqref{eq:inflaton-potential}, since its field dependence
is spurion suppressed, and is negligible over the
CMB-relevant field range.  See Ref.~\cite{Nakayama:2012dw}
for a supersymmetric new-inflation model in which a
supersymmetry-breaking Coleman--Weinberg correction can raise
the spectral index.}
\begin{equation}
V(s,a)=V_{\rm inf}(s)
+\frac{\lambda_a}{4}(a^2-f^2)^2,
\label{eq:full-potential}
\end{equation}
with
\begin{equation}
V_{\rm inf}(s)=V_0-Cs-\frac{m^2}{2}s^2
-\frac{\lambda_s}{4}s^4
+\frac{\lambda_6}{6\Lambda^2}s^6.
\label{eq:inflaton-potential}
\end{equation}
Here $C$, $m^2$, $\lambda_s$, and $\lambda_6$ are taken
to be positive. The even-power terms are consistent with
the discrete symmetries precisely because $S^2$ is a
singlet, while the linear term is generated by the small
real spurion $\mathcal C$. Eq.~\eqref{eq:full-potential}
should thus be regarded as a convenient benchmark
corresponding to a particular choice of the coefficients
of the symmetry-allowed operators. More general mixed
interactions deform the two valleys and lead to a
genuinely two-field inflationary trajectory, without
altering the basic existence of the two CP-conjugate
branches.

We focus on a quartic-dominated hilltop
regime in which the negative quartic term controls the
evolution away from the immediate vicinity of the
hilltop and the eventual end of slow roll.  The linear
and quadratic terms are treated as deformations of this
dynamics; in particular, the linear term can dominate the
slope sufficiently close to the origin and plays an
important role in determining the CMB spectral tilt~\cite{Takahashi:2013cxa}.  The
positive sextic term stabilizes the potential at large
field values and determines the vacuum at $s=v$.

We use the position $v$ of the  minimum as an
input and impose
\begin{equation}
 V_{\rm inf}'(v)=0,
 \qquad
 V_{\rm inf}(v)=0.
\end{equation}
These conditions determine the stabilizing coefficient
and the additive constant,
\begin{align}
 \frac{\lambda_6}{\Lambda^2}
 &=\frac{\lambda_s}{v^2}
 +\frac{m^2}{v^4}
 +\frac{C}{v^5},
\label{eq:lambda6-vacuum}\\
 V_0&=\frac{\lambda_s}{12}v^4
 +\frac{m^2}{3}v^2
 +\frac{5C}{6}v.
\label{eq:V0-vacuum}
\end{align}
Thus $\lambda_6$ is not an independent parameter in the
following analysis.  In the regime of interest, the
linear and quadratic terms are small deformations of the
quartic hilltop potential, and hence
\begin{equation}
 V_0\simeq\frac{\lambda_s}{12}v^4.
\label{eq:V0-leading}
\end{equation}
The inflationary energy scale is therefore determined
primarily by the quartic coupling and the vacuum position.

The inflaton mass around the minimum is
\begin{equation}
 m_\varphi^2=V_{\rm inf}''(v)
 =2\lambda_s v^2+4m^2+\frac{5C}{v}.
\label{eq:mphi}
\end{equation}
For the same small deformations,
\begin{equation}
m_\varphi^2\simeq2\lambda_s v^2.
\label{eq:mass}
\end{equation}
Thus both the vacuum energy and the inflaton mass are
controlled mainly by $\lambda_s$ and $v$.

\subsection{Quartic hilltop limit and the linear deformation}

To isolate the baseline hilltop dynamics, we first neglect the linear and
quadratic deformations.  For $s\ll v$, the potential then reduces to
\begin{equation}
 V_{\rm inf}(s)\simeq V_0-\frac{\lambda_s}{4}s^4.
\label{eq:quartic-limit}
\end{equation}
Using $|\eta_V|=1$ to estimate the end of slow roll gives
\begin{align}
 s_{\rm end}^2&\simeq\frac{V_0}{3\lambda_s\mpl^2},\\
 s_*^2&\simeq\frac{V_0}
 {2\lambda_s\mpl^2(N_*+3/2)},\\
 n_s&\simeq1-\frac{3}{N_*+3/2},
\label{eq:quartic-results}
\end{align}
where $\eta_V = \mpl^2 V_{\rm inf}''/V_{\rm inf}$, $M_{\rm Pl}=2.4\times10^{18}\ {\rm GeV}$ is the
reduced Planck mass,  and the subscript $*$ means that it is evaluated at the horizon exit of CMB scales.
For $N_*=40$ and $50$, this gives $n_s\simeq0.928$ and
$0.942$, respectively, substantially below the value preferred by the
recent ACT DR6 analysis~\cite{AtacamaCosmologyTelescope:2025blo} discussed below.  In the quartic limit, the
magnitude of the negative curvature sampled at the pivot scale is therefore
too large.

Restoring the deformations, the slope near the hilltop is
\begin{equation}
 -V_{\rm inf}'(s) \simeq C+m^2s+\lambda_ss^3,
\label{eq:slope}
\end{equation}
Although the linear term is negligible in the inflationary energy density, the
constant contribution $C$ to the slope can be important close to the
hilltop.  At fixed $N_*$, it moves the pivot value $s_*$ toward the origin,
where the quartic contribution $-3\lambda_ss_*^2$ to $V_{\rm inf}''$ is
smaller in magnitude, and thereby raises $n_s$~\cite{Takahashi:2013cxa}.
The mass term provides an additional, subleading modification of the
curvature.

We focus on the parameter region in which the linear term modifies the
motion near the CMB pivot, while the quartic term controls the subsequent
departure from the hilltop and the end of slow roll.

\subsection{Numerical evolution and CMB contours}

For the numerical analysis, we solve the homogeneous inflaton equation
 of motion.  Using the number of
$e$-folds, $N=\ln R$, as the time variable, where $R$ is the scale factor,
the background equations are
\begin{align}
 &s_{,NN}+(3-\epsilon_1)
 \left(s_{,N}+\mpl^2\frac{V_{\rm inf}'}{V_{\rm inf}}\right)=0,
\label{eq:inflaton-eom}\\
& \epsilon_1=\frac{s_{,N}^2}{2\mpl^2},\\
& H^2=\frac{V_{\rm inf}}{\mpl^2(3-\epsilon_1)}.
\label{eq:inflaton-H}
\end{align}
We define the end of inflation by $\epsilon_1=1$.  At the pivot scale,
located $N_*$ $e$-folds before  the end of slow roll, the scalar spectral index and
tensor-to-scalar ratio are evaluated, to first order in the Hubble-flow
parameters, as
\begin{equation}
 n_s=1-2\epsilon_1-\epsilon_2,
 \qquad
 r=16\epsilon_1
\label{eq:hubble-flow}
\end{equation}
with  $\epsilon_2=\frac{\dd\ln\epsilon_1}{\dd N}$.
The scalar amplitude is
\begin{equation}
 A_s=\frac{H_*^2}{8\pi^2\mpl^2\epsilon_{1*}}.
\label{eq:As}
\end{equation}

We fix
\begin{equation}
 v=10^{13}\ {\rm GeV},
 \qquad
 \Lambda=\mpl,
\end{equation}
and scan the $(C^{1/3},m)$ plane. This choice is motivated by the quality problem (see Eq.~(\ref{eq:quality})).
At each point, $\lambda_s$ is determined
by imposing $A_s=2.10\times10^{-9}$.  The vacuum conditions in
Eqs.~\eqref{eq:lambda6-vacuum} and \eqref{eq:V0-vacuum} then determine
$\lambda_6$ and $V_0$.

The ACT DR6 power spectra, combined with large-scale CMB information,
CMB-lensing data from ACT and Planck, and DESI Y1 baryon-acoustic-oscillation
data, give
\begin{equation}
 n_s=0.974\pm0.003
 \qquad (68\%~{\rm C.L.})
\label{eq:ACT-ns}
\end{equation}
within the six-parameter $\Lambda$CDM model~\cite{AtacamaCosmologyTelescope:2025blo}.  Since the
mapping between the pivot scale and the inflationary trajectory depends on
the post-inflationary expansion history, we show the results for
$N_*=40$.

\begin{figure*}[t]
\includegraphics[width=0.55\textwidth]{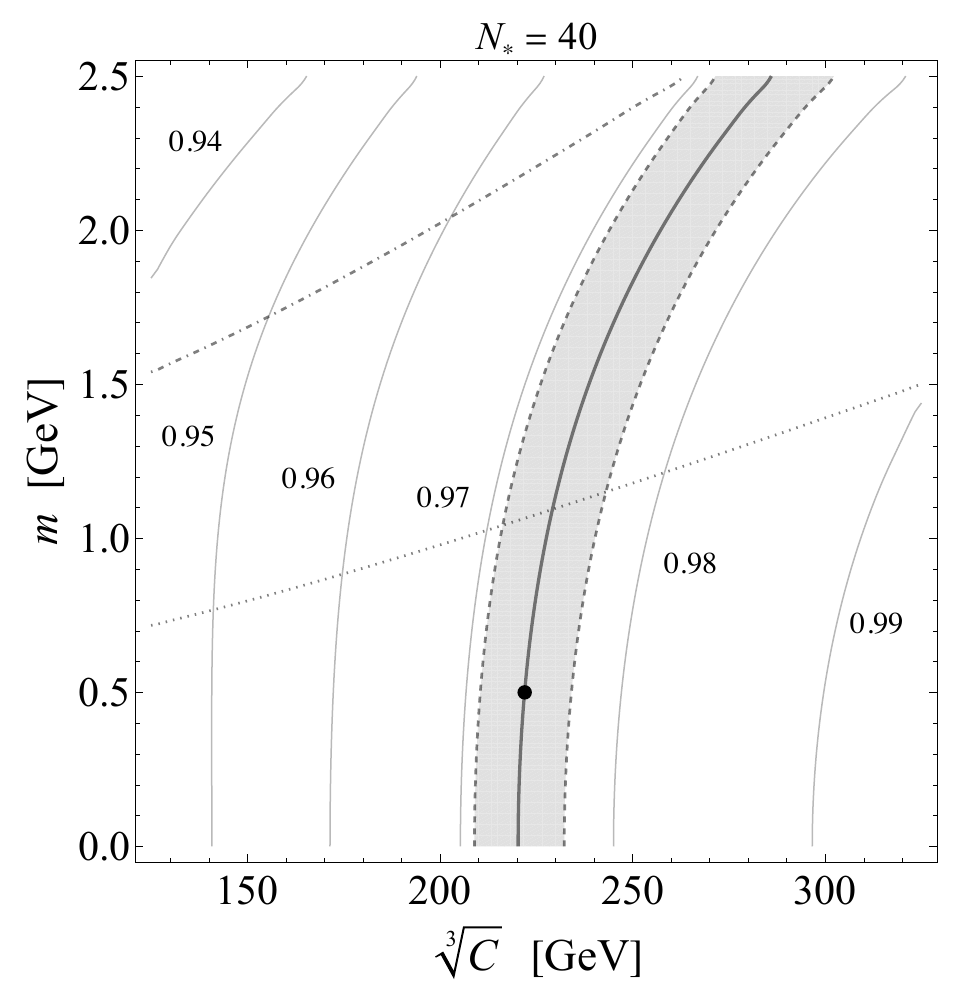}
 \caption{Scalar spectral index obtained from the numerical inflaton
 evolution for $N_*=40$, with
 $v=10^{13}\,$GeV and $\Lambda=\mpl$.   The solid thick 
 curve denotes the ACT DR6 central value $n_s=0.974$, and the thick dashed
 curves denote the boundaries of its $1\sigma$ interval,
 $n_s=0.971$ and $0.977$, and the $1\sigma$ allowed region is shaded with gray. 
 The dotted and dot-dashed lines represent the contour of the ratio of the mass to the Hubble parameter, $m_/H_* = 0.1$, and $0.2$, respectively.  The black dot marks the representative point in
 Table~\ref{tab:benchmark}.}
 \label{fig:scan}
\end{figure*}

Fig.~\ref{fig:scan} shows the resulting spectral index.  The ACT-preferred
region is obtained over a finite range of parameters.  Increasing the
linear slope shifts the pivot closer to the origin and gives the dominant
upward change in $n_s$, while the mass term produces a milder displacement
of the contours.

A representative ACT-compatible point is shown in
Table~\ref{tab:benchmark}.  The quoted precision is intended only to
indicate the relevant scales.

\begin{table}[t]
\caption{Representative CMB-normalized point for
$v=10^{13}\,$GeV and $\Lambda=\mpl$.}
\label{tab:benchmark}
\begin{ruledtabular}
\begin{tabular}{lc}
Quantity & Value\\
\hline
$N_*$ & $40$\\
$C^{1/3}$ & $2.22\times10^2\ {\rm GeV}$\\
$m$ & $5.0\times10^{-1}\ {\rm GeV}$\\
$\lambda_s$ & $2.367\times10^{-12}$\\
$n_s$ & $0.974$\\
$H_*$ & $10.5 \ {\rm GeV}$\\
$V_0^{1/4}$ & $6.66\times10^9\ {\rm GeV}$\\
$m_\varphi$ & $2.18\times10^7\ {\rm GeV}$
\end{tabular}
\end{ruledtabular}
\end{table}

\section{Post-inflationary CP non-restoration}
\label{sec:nonrestoration}

The two inflationary valleys of Eq.~\eqref{eq:full-potential} are
\begin{equation}
 a=+f,
 \qquad
 a=-f,
\label{eq:valleys}
\end{equation}
and are exchanged by CP.  Every continuous path from one branch to the other crosses the CP-fixed locus $a=0$.  Since the true vacuum is normalized to zero, the minimum potential energy on this locus is
\begin{equation}
 V_{\rm CP}\equiv\min_sV(s,0)=\frac{\lambda_af^4}{4}.
\label{eq:VCP}
\end{equation}
The homogeneous scalar energy satisfies
\begin{equation}
 \rho_S=\frac12\dot s^2+\frac12\dot a^2+V(s,a).
\label{eq:energy-monotonic}
\end{equation}
The largest energy available to a trajectory emerging from the hilltop is of order $V_0$.  A sufficient condition excluding a homogeneous transition to the conjugate branch is therefore
\begin{equation}
 \boxed{\frac{\lambda_af^4}{4}>V_0.}
\label{eq:barrier-condition}
\end{equation}
For $f=v$, Eq.~\eqref{eq:V0-vacuum} gives
\begin{equation}
 \lambda_a>
 \frac{\lambda_s}{3}
 +\frac{4m^2}{3v^2}
 +\frac{10C}{3v^3}.
\label{eq:lambda-a-bound}
\end{equation}
The right-hand side is of order $10^{-13}$ in the CMB region.  Hence a generic perturbative value $\lambda_a=\Order(1)$ places the CP-invariant ridge many orders of magnitude above the inflationary plateau.  Thus, the CMB normalization fixes the inflationary energy scale
well below the CP barrier, preventing the post-inflationary
trajectory from reaching the opposite branch.

For the separable potential in Eq.~\eqref{eq:full-potential},
$a=\pm f$ remains an exact solution independently of the
post-inflationary motion of $s$.  Fluctuations around either
branch have the constant positive mass
\begin{equation}
 m_a^2=2\lambda_a f^2,
\end{equation}
and therefore experience neither a tachyonic instability nor
parametric excitation induced by the inflaton oscillation.  For
$\lambda_a=\Order(1)$ and $f\sim v$, one has
$m_a\gg m_\varphi,H$, so the CP-odd mode remains decoupled
throughout inflation and reheating.

Thermal and quantum transitions between the two branches are
also negligible in the parameter region of interest.  Thermal
restoration of CP would require a temperature of order $f$, whereas
the reheating temperatures considered below satisfy $T_R\ll f$.
During inflation, transverse fluctuations are strongly suppressed
by $m_a\gg H_{\rm inf}$, while quantum tunneling across the ridge is
exponentially suppressed because
$\lambda_a f^4/4\gg H_{\rm inf}^4$.

Mixed $s$--$a$ interactions generically curve the inflationary
valley and induce a turning trajectory, an $s$-dependent transverse
mass, and possible nonadiabatic excitation of transverse fluctuations.
A detailed treatment of such effects generally requires a two-field
analysis.  Nevertheless, the absence of a homogeneous transition to
the CP-conjugate branch follows from the more general energetic
condition
\begin{equation}
 \min_s V(s,0)>\rho_{S,\rm ini}.
\end{equation}
This conclusion does not rely on the valley being exactly straight.
As long as inflation takes place on a sufficiently shallow plateau
whose energy lies well below the lowest point of the CP-invariant
ridge, a curved post-inflationary trajectory cannot reach the
conjugate branch.  Since Hubble expansion further decreases the total
scalar energy, this inequality is preserved during the subsequent
evolution.  The protection of the selected CP branch is therefore
robust against sizable deformations of the separable potential,
provided that they do not substantially lower the CP-invariant ridge.

\section{Reheating and nonthermal leptogenesis}
\label{sec:reheating}

Identifying the Nelson--Barr scalar with the inflaton
makes the interactions responsible for transmitting
spontaneous CP violation directly relevant to reheating.
We first discuss the visible decay channel induced by the
vector-like-quark threshold and then consider the Higgs
and right-handed-neutrino portals.
\subsection{Reheating through the vector-like-quark threshold}

Let
\begin{equation}
 \varphi=s-v
\end{equation}
denote the inflaton fluctuation about the vacuum.  Neglecting
the ordinary down-quark masses, the heavy vector-like quark
has the field-dependent mass
\begin{equation}
 m_D^2(\varphi)
 =
 \widehat M^2+B(\varphi)B^\dagger(\varphi).
\label{eq:mD-field}
\end{equation}
We define
\begin{equation}
 b_i\equiv
 \left.
 \frac{\partial B_i}{\partial\varphi}
 \right|_{\varphi=0}.
\end{equation}
Because CP is spontaneously broken in the vacuum, the
diagonal coupling of $\varphi$ to the heavy mass eigenstate
$D_H$ is not necessarily purely scalar.  It can be written as
\begin{equation}
 {\cal L}
 \supset
 -m_D\varphi\,
 \overline D_H
 \left(
   \kappa_S+i\kappa_P\gamma_5
 \right)D_H,
\label{eq:inflaton-heavy-coupling}
\end{equation}
where
\begin{align}
 \kappa_S
 &\equiv
 \frac{\operatorname{Re}(bB^\dagger)}{m_D^2}
 =
 \left.
 \frac{\partial\ln m_D}{\partial\varphi}
 \right|_{\varphi=0}
 \equiv
 \frac{c_S}{v},
 \label{eq:kappaD-scalar}
 \\
 \kappa_P
 &\equiv
 \frac{\operatorname{Im}(bB^\dagger)}{m_D^2}
 \equiv
 \frac{c_P}{v}.
 \label{eq:kappaD-pseudo}
\end{align}

For the vacuum parameters introduced in Sec.~II,
\begin{align}
 \kappa_S
 &=
 \frac{\eps^{2k}v}{2m_D^2}
 \sum_i(g_i+g_i')^2,
 \label{eq:kappaD-scalar-explicit}
 \\
 \kappa_P
 &=
 -\sigma\,
 \frac{\eps^{2k}f}{2m_D^2}
 \sum_i(g_i^2-g_i'^2).
 \label{eq:kappaD-pseudo-explicit}
\end{align}
Thus, even though $\varphi$ is the fluctuation of the
CP-even field $s$, a pseudoscalar coupling is generically
induced after expanding around a CP-breaking vacuum.
For $|B|\sim\widehat M$, $f\sim v$, and generic flavor
coefficients, $|c_S|$ and $|c_P|$ can naturally be of order unity, although either coefficient can be
suppressed by a special flavor structure.  In particular,
$c_P$ vanishes if
$\sum_i g_i^2=\sum_i g_i'^2$.

For $\eps^k\sim10^{-5}$ near the upper
limit suggested by Nelson--Barr quality and
$M=\Order(v)$, the representative parameter region satisfies
$m_\varphi<m_D$.  Decays containing an on-shell vector-like
quark are therefore kinematically forbidden in this case.\footnote{
 Since $m_D\propto\eps^k$, the tree-level
channels $\varphi\to D\bar d+\bar Dd,\,D \bar D$ open for
smaller $\varepsilon$; their widths are given in
Appendix~\ref{app:heavy-quark-decay}.
}  The diagonal
inflaton couplings to two light down-type quarks are
proportional to the corresponding light-quark masses, so their
two-body decay rates are strongly suppressed.

Integrating out $D_H$ in the limit $m_\varphi\ll 2m_D$ gives
the effective interactions~\cite{Inami:1982xt,Djouadi:2005gi}\footnote{
Although the phase of $S$ does not induce a strong CP phase, $\varphi$ can have an effective coupling to $G\widetilde G$ at finite momentum.  The heavy- and light-quark contributions cancel in the zero-momentum limit, as required by the absence of an induced $\bar\theta$.  For $m_d\ll m_\varphi\ll m_D$, however, their loop form factors are different, and the cancellation does not persist in the physical $\varphi\to gg$ amplitude.
}
\begin{equation}
 {\cal L}_{\varphi gg}
 =
 \frac{\alpha_s}{12\pi}
 \kappa_S\varphi
 G^a_{\mu\nu}G^{a\mu\nu}
 +
 \frac{\alpha_s}{8\pi}
 \kappa_P\varphi
 G^a_{\mu\nu}\widetilde G^{a\mu\nu}.
\label{eq:phi-gg}
\end{equation}
The first term follows from the field dependence of the
absolute value of the heavy mass, while the second follows
from the anomalous chiral rotation associated with its
field-dependent phase.

At leading order in $\alpha_s$ and $m_\varphi^2/m_D^2$,
the corresponding partial width is
\begin{align}
 \Gamma_{gg}
 &=
 \frac{\alpha_s^2m_\varphi^3}{72\pi^3}
 \left[
   \kappa_S^2
   +\frac{9}{4}\kappa_P^2
 \right]
 \nonumber\\
 &=
 \frac{\alpha_s^2c_g^2}{72\pi^3}
 \frac{m_\varphi^3}{v^2},
\label{eq:Gamma-gg}
\end{align}
where
\begin{equation}
 c_g^2
 \equiv
 c_S^2+\frac{9}{4}c_P^2.
\label{eq:cg-def}
\end{equation}

Evaluating $\alpha_s$ at a renormalization scale of order
$m_\varphi$, the leading-order width can be displayed as
\begin{align}
 \Gamma_{gg}
 \simeq{}&
 4.9\times10^{-12}\ {\rm GeV} \,c_g^2
 \left(
   \frac{\alpha_s(m_\varphi)}{0.033}
 \right)^2 \nonumber\\
 &\times
 \left(
   \frac{m_\varphi}{10^{7}\ {\rm GeV}}
 \right)^3
 \left(
   \frac{10^{13}\ {\rm GeV}}{v}
 \right)^2 .
\label{eq:Gamma-gg-numerical}
\end{align}
The
reheating temperature in the sudden-decay approximation is
\begin{equation}
 T_R
 =
 \left(
   \frac{90}{\pi^2g_*}
 \right)^{1/4}
 \sqrt{\Gamma_{\rm tot}M_{\rm Pl}},
\label{eq:TR-gluon}
\end{equation}
where $\Gamma_{\rm tot}$ is the total decay rate. 
If the decay into gluons dominates the total inflaton decay rate, we have
\begin{align}
 T_R
 \simeq{}&
 1.9\times10^3\ {\rm GeV}\,c_g
 \left(
   \frac{106.75}{g_*}
 \right)^{1/4}
 \left(
   \frac{\alpha_s(m_\varphi)}{0.033}
 \right)
\nonumber\\
 &\times
 \left(
   \frac{m_\varphi}{10^{7}\ {\rm GeV}}
 \right)^{3/2}
 \left(
   \frac{10^{13}\ {\rm GeV}}{v}
 \right).
\label{eq:TR-gluon-numerical}
\end{align}
Thus, for the representative inflationary point, the reheating temperature is  approximately
$10^3\, {\rm GeV}$.

\subsection{Higgs portal and electroweak naturalness}

The discrete symmetries also allow the Higgs portal, and
expanding about the vacuum gives the trilinear
interaction
\begin{equation}
 \Delta V
 \supset
 \lambda_{SH}v\varphi|H|^2.
\end{equation}
For $m_\varphi$ well above the electroweak scale, the decay rate into a pair of Higgs particles is
\begin{equation}
 \Gamma_{HH}
 \simeq
 \frac{\lambda_{SH}^2v^2}{8\pi m_\varphi}.
\label{eq:Gamma-HH}
\end{equation}
This channel dominates over Eq.~\eqref{eq:Gamma-gg} for
\begin{equation}
 |\lambda_{SH}|
 \gtrsim
 \frac{\alpha_s|c_g|}{3\pi}
 \left(\frac{m_\varphi}{v}\right)^2
 =\Order(10^{-14}),
\label{eq:portal-critical}
\end{equation}
where we used $m_\varphi \sim 10^{7}\,$GeV, $v \sim 10^{13}$\,GeV and $c_D \sim 1$ in the last equality.

The same portal, however, contributes to the Higgs mass
parameter,
\begin{equation}
 \delta m_H^2
 \sim
 \lambda_{SH} v^2.
\end{equation}
In the absence of an additional mechanism protecting the
electroweak scale, avoiding a cancellation in the Higgs
mass parameter suggests
\begin{equation}
 |\lambda_{SH}|
 \lesssim
 \Order\left(\frac{m_h^2}{v^2}\right)
 \sim10^{-22}
\label{eq:portal-natural}
\end{equation}
for $v\sim10^{13}\ {\rm GeV}$.  Thus, the decay through the Higgs portal is subdominant unless the associated fine-tuning is accepted. 

\subsection{Right-handed neutrinos and nonthermal leptogenesis}

The spontaneous CP phase of the Nelson--Barr scalar can
also be communicated to the lepton sector.  We introduce
right-handed neutrinos $N_i$ through~\cite{Branco:2003rt}
\begin{equation}
 -\mathcal L_N
 =
 \frac12
 \left(\xi_{ij}S+\xi'_{ij}S^*\right)
 \overline{N_i^c}N_j
 +(y_\nu)_{\alpha i}
 \overline L_\alpha\widetilde H N_i
 +\mathrm{h.c.},
\label{eq:LN}
\end{equation}
where $\xi$ and $\xi'$ are real symmetric matrices,
$y_\nu$ is real before spontaneous CP breaking, and
$\widetilde H=i\sigma_2H^*$.

In the vacuum labeled by $\sigma$, the Majorana mass
matrix is
\begin{equation}
 \mathcal M_N^{(\sigma)}
 =
 \frac{1}{\sqrt2}
 \left[
  (\xi+\xi')v
  +i\sigma(\xi-\xi')f
 \right].
\label{eq:MN-vacuum}
\end{equation}
We define its Takagi factorization by
\begin{equation}
 U_N^T\mathcal M_N^{(\sigma)}U_N
 =
 \widehat M_N
 \equiv
 \operatorname{diag}(M_1,M_2,M_3)
\label{eq:Takagi}
\end{equation}
with $0<M_1<M_2<M_3$.
In the heavy-neutrino mass basis,
\begin{equation}
 Y_\nu=y_\nu U_N,
 \qquad
 m_\nu
 =
 -v_H^2Y_\nu\widehat M_N^{-1}Y_\nu^T.
\label{eq:seesaw}
\end{equation}
Although all fundamental couplings are real, $U_N$ is
generically complex.  The spontaneous phase of $S$ can
therefore enter both the low-energy neutrino mass matrix
and the combinations
$\operatorname{Im}[(Y_\nu^\dagger Y_\nu)_{1j}^2]$
responsible for leptogenesis.

Since the Nelson--Barr scalar is identified with the
inflaton, the same interaction also produces the heavy
neutrinos during reheating.  Its
coupling in the heavy-neutrino mass basis is
\begin{equation}
 \widehat y^\varphi
 =
 \frac{1}{\sqrt2}
 U_N^T(\xi+\xi')U_N,
\label{eq:yphi}
\end{equation}
and the diagonal decay width is
\begin{equation}
 \Gamma(\varphi\to N_iN_i)
 =
 \frac{|\widehat y^\varphi_{ii}|^2}{32\pi}
 m_\varphi
 \left(
  1-\frac{4M_i^2}{m_\varphi^2}
 \right)^{3/2}.
\label{eq:Gamma-N}
\end{equation}

We consider the nonthermal regime
\begin{equation}
 2M_1< m_\varphi < M_1+M_2,
 \qquad
 T_R\ll M_1,
\label{eq:nonthermal-conditions}
\end{equation}
in which $N_1$ is produced directly by inflaton decay,
while its thermal production and inverse-decay washout are
suppressed.  Once the spontaneous CP phase has been
transmitted to $Y_\nu$, the subsequent mechanism is the
standard nonthermal leptogenesis from inflaton
decay~\cite{Lazarides:1990huy,Asaka:1999jb,Asaka:1999yd}.

We denote the total CP asymmetry in $N_1$ decay by
\begin{equation}
 \epsilon_{N_1}
 \equiv
 \frac{
  \Gamma(N_1\to LH)
  -\Gamma(N_1\to\overline L H^\dagger)
 }{
  \Gamma(N_1\to LH)
  +\Gamma(N_1\to\overline L H^\dagger)
 }.
\label{eq:epsilonN1-def}
\end{equation}
For a hierarchical heavy-neutrino spectrum,
$M_{2,3}\gg M_1$, it is given by
\begin{equation}
 \epsilon_{N_1}
 \simeq
 -\frac{3}{
  16\pi(Y_\nu^\dagger Y_\nu)_{11}}
 \sum_{j\ne1}
 \operatorname{Im}
 \left[
  (Y_\nu^\dagger Y_\nu)_{1j}^2
 \right]
 \frac{M_1}{M_j},
\label{eq:epsilonN1-hierarchical}
\end{equation}
and satisfies the Davidson--Ibarra bound
\begin{align}
 |\epsilon_{N_1}|
 &\lesssim
 \frac{3M_1}{16\pi v_H^2}(m_{\nu 3}-m_{\nu 1})
 \nonumber\\
 &\simeq
 1\times 10^{-9}
 \left(
  \frac{M_1}{10^7\ {\rm GeV}}
 \right)
 \nonumber\\
 &\hspace{2.0cm}\times
 \left(
  \frac{m_{\nu 3}}{0.05\ {\rm eV}}
 \right),
\label{eq:DI-bound}
\end{align}
where the numerical expression uses $v_H=174\ {\rm GeV}$
and a normal hierarchical light-neutrino spectrum~\cite{Davidson:2002qv}.

In the sudden-decay approximation, each inflaton decay
into $N_1N_1$ produces two heavy neutrinos, giving
\begin{equation}
 Y_{N_1}
 \equiv
 \frac{n_{N_1}}{s}
 =
 \frac32
 \Br_1\frac{T_R}{m_\varphi},
 \qquad
 \Br_1\equiv
 \Br(\varphi\to N_1N_1).
\label{eq:YN}
\end{equation}
Electroweak sphalerons then yield
\begin{align}
 |Y_B|
 &=
 \frac{28}{79}
 \kappa|\epsilon_{N_1}|Y_{N_1}
 \nonumber\\
 &=
 \frac{42}{79}
 \kappa\Br_1|\epsilon_{N_1}|
 \frac{T_R}{m_\varphi},
\label{eq:YB-nonthermal}
\end{align}
where $\kappa\leq1$ parametrizes residual washout and
subsequent entropy dilution.  In the regime
$T_R/M_1\ll1$, with prompt $N_1$ decay, one expects
$\kappa\simeq1$.

For an order-of-magnitude estimate, we assume that
the decay into right-handed neutrinos dominates the
inflaton width,
\begin{equation}
 \Gamma_{\rm tot}
 \simeq
 \Gamma(\varphi\to N_1N_1),
 \qquad
 \Br_1\simeq1.
\end{equation}
In the absence of an accidental cancellation in the
Majorana mass matrix, it is natural to parametrize the
inflaton coupling as
\begin{equation}
 |\widehat y^\varphi_{11}|
 =
 c_N\frac{M_1}{v},
 \qquad
 c_N=\mathcal O(1).
\label{eq:cN-def}
\end{equation}
The total inflaton width is then
\begin{equation}
 \Gamma_{\rm tot}
 \simeq
 \frac{c_N^2M_1^2}{32\pi v^2}
 m_\varphi
 \left(
  1-\frac{4M_1^2}{m_\varphi^2}
 \right)^{3/2}.
\label{eq:Gamma-N-dominated}
\end{equation}

For the low-scale benchmark in
Table~\ref{tab:benchmark}, we take
\begin{equation}
 m_\varphi\simeq2.2\times10^7\ {\rm GeV},
 \qquad
 M_1\simeq8\times10^6\ {\rm GeV}.
\label{eq:low-N-benchmark}
\end{equation}
Thus $M_1\simeq0.37m_\varphi$, close to the kinematic
threshold but without a severe phase-space suppression.
Eqs.~\eqref{eq:TR-gluon} and
\eqref{eq:Gamma-N-dominated} give
\begin{align}
 T_R
 \simeq{}&
 1.8\times10^5\ {\rm GeV}\,c_N
 \left(
  \frac{M_1}{8\times10^6\ {\rm GeV}}
 \right)
 \left(
  \frac{m_\varphi}{2.2\times10^7\ {\rm GeV}}
 \right)^{1/2}
 \nonumber\\
 &\times
 \left(
  \frac{10^{13}\ {\rm GeV}}{v}
 \right)
 \left[
  \frac{1-4M_1^2/m_\varphi^2}{0.47}
 \right]^{3/4}
 \left(
  \frac{106.75}{g_*}
 \right)^{1/4}.
\label{eq:TR-N-low}
\end{align}
This satisfies $T_R/M_1\simeq2\times10^{-2}c_N$, so
thermal production and washout remain suppressed.

Using the Davidson--Ibarra bound, the resulting baryon
asymmetry is bounded by
\begin{align}
 |Y_B|
 \lesssim{}&
 3.4\times10^{-12}\,
 \kappa c_N
 \left(
  \frac{M_1}{8\times10^6\ {\rm GeV}}
 \right)^2
 \left(
  \frac{2.2\times10^7\ {\rm GeV}}{m_\varphi}
 \right)^{1/2}
 \nonumber\\
 &\times
 \left(
  \frac{10^{13}\ {\rm GeV}}{v}
 \right)
 \left[
  \frac{1-4M_1^2/m_\varphi^2}{0.47}
 \right]^{3/4}
 \left(
  \frac{m_{\nu3}}{0.05\ {\rm eV}}
 \right).
\label{eq:YB-low-bound}
\end{align}
For order-one $c_N$ and $\kappa$, this is about a factor
of thirty below the observed value,
$Y_B^{\rm obs}\simeq8.7\times10^{-11}$.
The low-scale benchmark therefore does not generate a
sufficient baryon asymmetry through standard hierarchical
nonthermal leptogenesis.  This deficit could be compensated
by a resonant enhancement of the CP
asymmetry if two right-handed neutrinos are quasi-degenerate.
In the next section we instead pursue the possibility of raising the
inflaton and reheating scales. From Eqs.~(\ref{eq:YB-low-bound}) and (\ref{eq:mass}), we can see that the upper bound on $Y_B$ scales as $v^{1/2}$ so that we need to increase the inflaton vev at least by a factor of $10^3$.

\section{An extension with higher CP breaking scale}
\label{sec:highscale-spurion}

Let us here extend the model given in Table~\ref{tab:charges}
by replacing $X$ with the spurions $\beta_M$, $\beta_i$, and $\beta'_i$. To be concrete, we
  replace the common $X$ insertion in Eq.~\eqref{eq:LNB} by
\begin{align}
 -\mathcal L_q^{\rm(high)}={}&
 \beta_M M\,\overline D_LD_R
 +(\beta_iS+\beta_i'S^*)\overline D_Ld_{Ri}
 \nonumber\\
 &+y^d_{ij}H\overline Q_{Li}d_{Rj}
 +\mathrm{h.c.},
\label{eq:high-spurion-L}
\end{align}
where $\beta_M$, $\beta_i$, and $\beta_i'$ are dimensionless spurions. Their discrete charges are listed
in Table~\ref{tab:high-spurion-charges}.
\begin{table}[t]
\caption{Charges of the small spurions and $d_{Ri}$.}
\label{tab:high-spurion-charges}
\begin{ruledtabular}
\begin{tabular}{c|cccc}
 & $\beta_M$ & $\beta_i$ & $\beta_i'$ &$d_{Ri}$\\
\hline
$\Zfour$
 & $0$ & $0$ & $0$ &$0$ \\
$\ZfourN$
 & $k$ & $-k$ & $-k$ &$k$\\
 U(3)$_{d_R}$ & $\bm 1$ & $\bm{\bar{3}}$ & $\bm{\bar{3}}$ & $\bm{{3}}$
\end{tabular}
\end{ruledtabular}
\end{table}
Their background values are taken to be real and of a
common order~(see Eq.~(\ref{eq:epsilon})),
\begin{equation}
 \beta_M\sim \beta\sim \beta'
 \equiv\delta\lesssim10^{-5}.
\label{eq:small-spurion-backgrounds}
\end{equation}
The real backgrounds therefore do not introduce an
additional CP phase.
Here $\beta$ and $\beta'$ represent the modulus of  $\beta_i$ and $\beta'_i$, respectively. 
The flavor vectors $\beta_i$ and $\beta'_i$ are taken to transform as anti-triplet spurions of the right-handed down-quark flavor symmetry.\footnote{The Standard-Model Yukawa matrices may also be treated
as the quark-flavor spurions.  Their insertion only
dresses the flavor contractions considered below and does
not modify the power counting in the small spurions
$\beta_M$, $\beta$, and $\beta'$.} We also impose an approximate interchange symmetry
\begin{equation}
 \mathcal R:\qquad
 S\leftrightarrow S^*,
 \qquad
 \beta_i\leftrightarrow\beta'_i.
\label{eq:spurion-reflection}
\end{equation}
The real spurion backgrounds preserve CP, while $\beta_i\neq \beta'_i$ breaks the interchange $\mathcal R$ symmetry and allows the two independent flavor vectors needed for an order-one CKM phase. Because the new spurions enter the quark sector, the scalar potential assumed in Eqs.~\eqref{eq:full-potential} and \eqref{eq:inflaton-potential}, including its separable form, is unchanged at tree level.

A pure scalar or gluonic operator must be a flavor singlet. The lowest $\mathcal R$-odd singlet is therefore
\begin{equation}
\beta_i \beta_i-\beta'_i \beta'_i
 =\mathcal O(\delta^2),
\end{equation}
where the sum over $i$ is understood.
The leading direct correction is then
\begin{align}
 \Delta\mathcal L_\theta^{\rm(high)}
 &\sim
 \mathcal (\beta_i \beta_i-\beta'_i \beta'_i)\,
 \frac{\operatorname{Im}(S^2)}{\Lambda^2}
 \nonumber\\
 &\quad\times
 \frac{\alpha_s}{8\pi}
 G^a_{\mu\nu}\widetilde G^{a\mu\nu}.
\end{align}
For $M\sim v\sim f$, an extra contribution to the strong CP phase is
\begin{align}
 \delta\bar\theta
 &\sim\delta^2\frac{v^2}{\Lambda^2}.
\label{eq:high-spurion-quality}
\end{align}
which is safely below the experimental limit, e.g. for  $\delta=10^{-5}$, $v=f=10^{16}\,$GeV, and $\Lambda = \mpl$.
Similarly,  complex corrections to the vector-like mass and the upper-right quark-mass block are also suppressed  by the same two-spurion invariant, and their contributions to the strong CP phase can be suppressed. 
 
 The right-handed-neutrino matrices are treated analogously as fundamental flavor spurions; for the couplings $|\xi_{ij}|,|\xi'_{ij}|\lesssim 10^{-5}$, their contributions to the same $\mathcal R$-odd invariants are at most comparable to the quark-sector estimates. Also, the lightest right-handed neutrino mass can become comparable to the inflaton mass. 

At $v\sim10^{16}\,$GeV the inflaton mass at the potential minimum becomes as large as $m_\varphi=\mathcal O(10^{10})\,$GeV, and the reheating temperature can be as high as 
$T_R\simeq10^7\ {\rm GeV}$,
for which nonthermal leptogeneis is possible.
The high-scale completion therefore enables nonthermal leptogenesis while leaving the inflationary valley and the domain-wall argument intact.

\section{Discussion and conclusions}
\label{sec:conclusion}

We have proposed that the Nelson--Barr scalar
itself drives hilltop inflation.  A CP-symmetric
double-well potential for the imaginary component of the
complex scalar gives two CP-conjugate valleys, and
inflation proceeds within one of them.  The accelerated
expansion homogenizes the spontaneous CP phase and
dilutes any preexisting domain walls.

The selected branch also remains protected after
inflation.  Any continuous trajectory connecting the two
valleys must cross the CP-invariant ridge.  Once the CMB
normalization is imposed, the coupling controlling the
hilltop energy $V_0$ is very small, whereas the ridge height $V_{\rm CP}$ is
set by $\lambda_af^4$.  For $f\sim v$ and an ordinary
perturbative value $\lambda_a=\Order(1)$, one therefore
finds
\begin{equation}
 V_{\rm CP}\gg V_0,
\end{equation}
so the post-inflationary trajectory cannot reach the
opposite branch.

Near the CMB region, the inflaton dynamics is described
by quartic hilltop inflation deformed by small linear and
quadratic terms.  The undeformed quartic limit predicts
a spectral index below the ACT-preferred range.  The
CP-preserving linear term changes the small slope near
the hilltop, moves the pivot value toward the origin, and
reduces the magnitude of the negative quartic curvature~\cite{Takahashi:2013cxa}.
Numerical integration of the inflaton evolution exhibits
regions consistent with
$n_s=0.974\pm0.003$~\cite{AtacamaCosmologyTelescope:2025blo}.  

For the separable potential, the two CP-breaking
valleys are straight and the transverse mode has a
positive, field-independent mass.  More general mixed
interactions curve the trajectory and introduce an
inflaton-dependent transverse mass, requiring a genuine
two-field analysis of turning and nonadiabatic
fluctuations.  Nevertheless, transitions
between the two CP-conjugate branches remain generically
forbidden as long as the inflationary plateau lies well
below the lowest point of the CP-invariant ridge.  The
post-inflationary scalar system then simply does not have
enough energy to reach the ridge, irrespective of the
curvature of its trajectory.  This conclusion is therefore
robust against generic mixed interactions that do not
substantially lower the CP-invariant ridge.

The Nelson--Barr interactions also connect inflation to
the subsequent thermal history.  When kinematically allowed,
they induce tree-level inflaton decays into a vector-like quark
and a light down-type quark, as well as into a pair of
vector-like quarks.  When these channels are closed, the
field-dependent vector-like-quark threshold generically induces
a loop-level decay into gluons. A Higgs
portal can provide more efficient reheating, although a
large portal entails a severe electroweak-scale tuning.
Couplings to right-handed neutrinos are especially
well-motivated: they transmit the spontaneous CP phase
to the light-neutrino sector and simultaneously allow
the inflaton to produce heavy Majorana neutrinos
nonthermally.  Their subsequent decays can account for
the baryon asymmetry in a model where the CP breaking scale is allowed to be as high as $10^{16}\,$GeV.

The strong CP problem and the origins of the CKM and
PMNS phases are closely related questions concerning CP
in the fundamental theory.  In the present scenario, the
Nelson--Barr sector not only communicates spontaneous CP
violation to the Standard Model, but also drives
inflation, reheats the visible Universe, enables
baryogenesis, and naturally removes the domain-wall
problem associated with its own symmetry breaking.

 \appendix
 \section{Inflaton decays involving the vector-like quark}
\label{app:heavy-quark-decay}

The value $\eps^k\sim10^{-5}$ should be regarded as an
approximate upper limit from Nelson--Barr quality rather
than as a lower limit~\cite{Murai:2024alz}.  Smaller
values further suppress the quality corrections without
reducing the CKM phase, since both $\widehat M$ and $B_i$
are proportional to the same factor $\eps^k$.  They do,
however, lower the physical vector-like-quark mass,
\begin{equation}
 m_D=\eps^k\mu_D
 \end{equation}
 with
 \begin{equation}
 \mu_D^2=
 M^2+\frac12\sum_i
 \left[
  (g_i+g_i')^2v^2+(g_i-g_i')^2f^2
 \right].
\label{eq:mD-epsilon}
\end{equation}
Consequently, 
for $m_\varphi>m_D$,
the decay channel $\varphi \to D \bar d +  \bar D d$ opens.

The decay width into one heavy and one light quark is
therefore
\begin{align}
 \Gamma_{D d}
 &\equiv
 \sum_{\alpha=1}^{3}
 \left[
  \Gamma(\varphi\to D\bar d_\alpha)
  +\Gamma(\varphi\to\bar Dd_\alpha)
 \right]
 \\
 &=
 \frac{3m_\varphi}{8\pi}
 \left(
  b_i b_i^*-
  \frac{|b_i B_i^*|^2}{m_D^2}
 \right)
 \left(1-\frac{m_D^2}{m_\varphi^2}\right)^2,
\label{eq:Gamma-Dd}
\end{align}
where summation over $i$ is understood.
This tree-level channel generically dominates over the
loop-induced gluon mode once it is open.

For completeness, if $m_\varphi>2m_D$, the decay into a pair of the heavy quarks is allowed. We obtain
\begin{equation}
 \Gamma(\varphi\to D\bar D)
 =
 \frac{3m_\varphi}{8\pi m_D^2}
 \left[
  (\operatorname{Re}[b_i B_i^*])^2\beta^3
  +(\operatorname{Im}[b_i B_i^*])^2\beta
 \right]
 \end{equation}
 with
 \begin{equation}
     \beta \equiv
 \sqrt{1-\frac{4m_D^2}{m_\varphi^2}}.
\label{eq:Gamma-DD}
\end{equation}

The produced heavy quarks quickly decay into the SM particles, and therefore they do not pose any cosmological problem.

%%%%%%%%%%%%%%%%%%%%%%%%%%%%%%%%%%%%%
\section*{Acknowledgments}
%%%%%%%%%%%%%%%%%%%%%%%%%%%%%%%%%%%%%
The author is grateful to Kazunori Nakayama and Kai Murai for useful comments on the manuscript.
This work was supported by JSPS KAKENHI Grant Numbers 25H02165 and 26K00695. This work was also supported by the World Premier International Research Center Initiative (WPI), MEXT, Japan, and by COST Action COSMIC WISPers CA21106, supported by COST (European Cooperation in Science and Technology). 

%%%%%%%%%%%%%%% References %%%%%%%%%%%%%%%%
\bibliographystyle{apsrev4-1}
\bibliography{ref}
%%%%%%%%%%%%%%%%%%%%%%%%%%%%%%%%%%%%%%%%%%%
\end{document}